\documentclass[aps,twocolumn,superscriptaddress,amsmath,amssymb]{revtex4-2}
\usepackage{graphicx}
\usepackage{dcolumn}
\usepackage{bm}
\usepackage{braket}
\usepackage{appendix}
\usepackage{subfigure}
\usepackage{hyperref}
\usepackage{color}
\usepackage{xcolor}
\usepackage{amsmath}
\usepackage{nccmath}
\usepackage[english]{babel}
\usepackage{svg}
\usepackage{graphicx,float}

\begin{document}
\title[2D nanoparticle array]{On-demand assembly of optically-levitated nanoparticle arrays in vacuum}

\author{Jiangwei Yan}
\email[These authors contributed equally to this work]{}
\affiliation{The State Key Laboratory of Quantum Optics and Quantum Optics Devices, Institute of Opto-Electronics, Shanxi University, Taiyuan 030006, China}
\affiliation{Collaborative Innovation Center of Extreme Optics, Shanxi University, Taiyuan, Shanxi 030006, China}

\author{Xudong Yu}
\email[These authors contributed equally to this work]{}
\affiliation{The State Key Laboratory of Quantum Optics and Quantum Optics Devices, Institute of Opto-Electronics, Shanxi University, Taiyuan 030006, China}
\affiliation{Collaborative Innovation Center of Extreme Optics, Shanxi University, Taiyuan, Shanxi 030006, China}

\author{Zheng Vitto Han}
\affiliation{The State Key Laboratory of Quantum Optics and Quantum Optics Devices, Institute of Opto-Electronics, Shanxi University, Taiyuan 030006, China}
\affiliation{Collaborative Innovation Center of Extreme Optics, Shanxi University, Taiyuan, Shanxi 030006, China}

\author{Tongcang Li}
\affiliation{Department of Physics and Astronomy, Purdue University, West Lafayette, IN, USA}

\author{Jing Zhang}
\email{Email: jzhang74@sxu.edu.cn, jzhang74@yahoo.com}
\affiliation{The State Key Laboratory of Quantum Optics and Quantum Optics Devices, Institute of Opto-Electronics, Shanxi University, Taiyuan 030006, China}
\affiliation{Collaborative Innovation Center of Extreme Optics, Shanxi University, Taiyuan, Shanxi 030006, China}

\date{\today}

\begin{abstract}
{\normalsize
Realizing a large-scale fully controllable quantum system is a challenging task in current physical research and has broad applications. Ultracold atom and molecule arrays in optical tweezers in vacuum have been used for quantum simulation, quantum metrology and quantum computing~\cite{Antoine30,Lukin31,Matthew2019,Kaufman32,Anderegg2019}. Recently, quantum ground state cooling of the center-of-mass motion of a single optically levitated nanoparticle in vacuum was demonstrated~\cite{Markus14,Markus15,Lukas16}, providing unprecedented opportunities for studying macroscopic quantum mechanics~\cite{Vitali11,Zhang12,Hendrik13,Oriol20} and precision measurements~\cite{John7,Carney_2021}. In this work, we create a reconfigurable optically-levitated nanoparticle array in vacuum. Our optically-levitated nanoparticle array allows full control of individual nanoparticles to form an arbitrary pattern and detect their motion. As a concrete example, we choose two nanoparticles without rotation signals from an array to synthesize a nanodumbbell in-situ by merging them into one trap. The nanodumbbell synthesized in-situ can rotate beyond 1 GHz. Our work provides a new platform for studying macroscopic many-body physics~\cite{liu2020,Kishan38,Svak42,Rieser2022,Vijayan2022} and quantum sensing~\cite{John7,Carney_2021}.}
\end{abstract}

\maketitle

Optical levitation employs forces exerted by strongly focused light fields to capture and manipulate microparticles and nanoparticles~\cite{Tongcang3,Millen4,Quidant5}. As a result, an optically levitated system is naturally isolated from environmental  disturbances. In particular,  it possess  extremely low damping in high vacuum and thus has an ultrahigh mechanical quality factor as an excellent optomechanical system. Hence, this system has attracted abroad attention and provides a powerful platform for precision measurements~\cite{John7,Carney_2021} and fundamental physics investigations~\cite{Vitali11,Zhang12,Hendrik13,Oriol20}. The center-of-mass (CoM) motion  of an optically levitated nanoparticle has been cooled to the quantum ground state~\cite{Markus14,Markus15,Lukas16}, and can  be used to generate non-Gaussian macroscopic quantum states in future. Besides CoM motion, libration~\cite{Li21},  rotation~\cite{Moore22}, and their coupling with internal degrees of freedoms (e.g., phonons, magnons, spin defects) of the levitated nanoparticle also provide rich physics to explore. In particular, the free rotation of a rigid body exhibits fascinating behaviors in both classical and quantum regimes due to its non-linear dynamics. We can now drive a single levitated nanoparticle to rotate at GHz frequencies~\cite{Lukas25,Li26,Li27,Jin28}, control its  rotation  with ultra-high precision~\cite{Kuhn23,Moore22}, and cool its librations to sub-kelvin temperatures~\cite{Bang2020,Lukas29}.

\begin{figure*}[t]
\centering
\centerline{\includegraphics[width=0.9\linewidth]{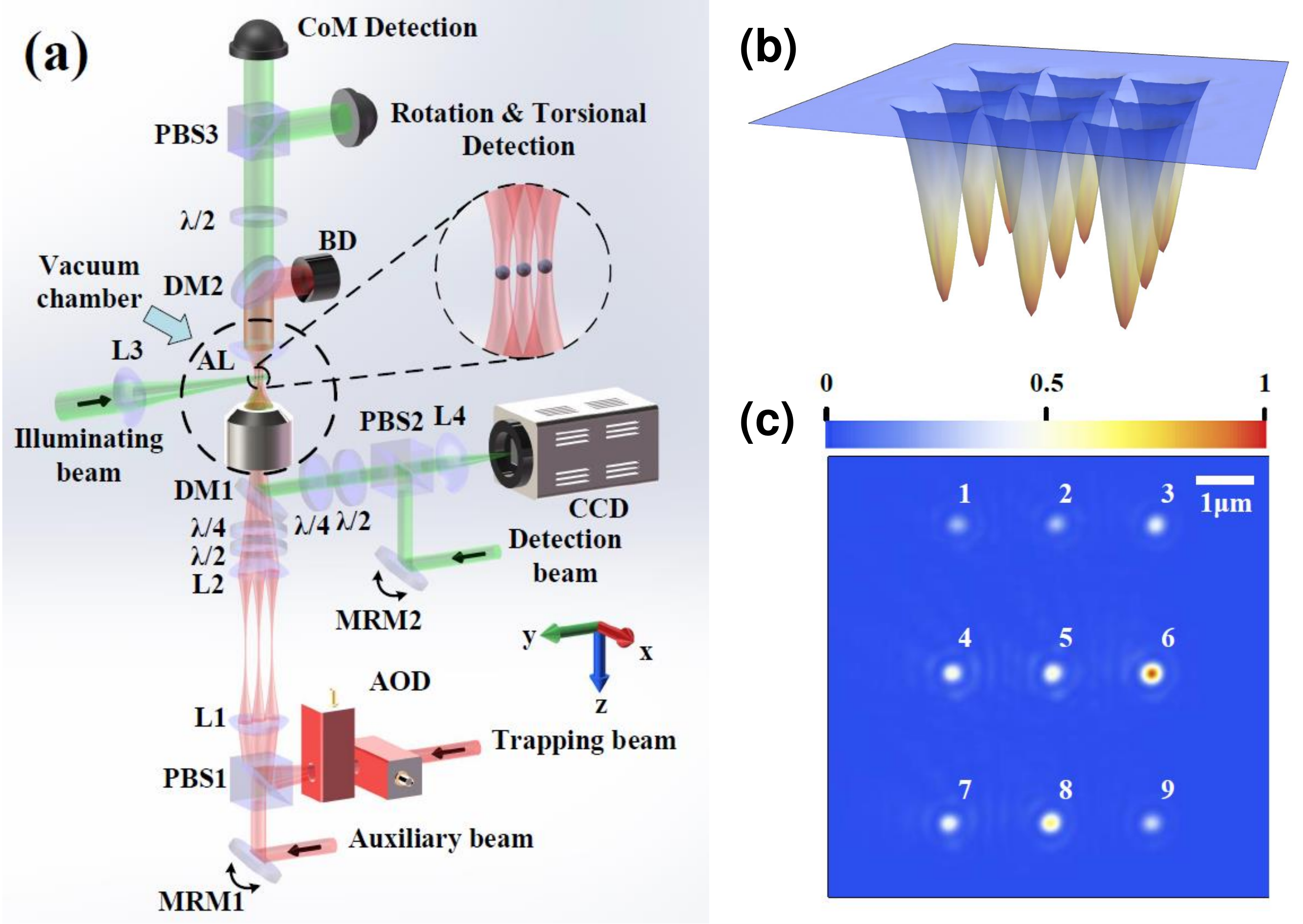}}
\caption{\textbf{Experimental setup and an image of an optically-levitated nanoparticle array.} (a) The 2D trap array is produced by passing a 1064 nm laser through a pair of orthogonal acousto-optic deflectors (AOD). The laser beams created by the AOD are imaged with a 4f image system onto a high NA (NA=0.95) objective lens, which creates an array of tightly focused optical tweezers in a vacuum chamber. The optical tweezers propagate against gravity. For particle assembly, we use an auxiliary moving tweezer at 1064 nm superimposed on the trap array  with a polarized beam splitter (PBS1). This auxiliary beam is deflected by a motor-driven reflective mirror (MRM1). For particle motion detection, we overlap a probe beam at 532 nm with the trapping beams using a dichroic mirror (DM1). This 532 nm beam is deflected by another motor-driven reflective mirror (MRM2) to measure the motion of an arbitrary particle in the array. The 1064 nm trap beams and the 532 nm probe beam after the high NA objective are collimated by an aspherical lens (AL) and seperated by another dichroic mirror (DM2) for detecting the motion of trapped nanoparticles. For particle imaging, an imaging beam at 532 nm is used to illuminate the particles orthogonally. The scattered light is collected by the same high NA objective to form an image on an charge-coupled device (CCD). $ \lambda/2$: half-wave plate; L1-4:spherical lens;  BD: beam dump. (b) A simulated trapping potential of a  $3 \times 3$ two-dimensional (2D) array of optical tweezers. (c) An image of a $3\times3 $  array of optically levitated nanoparticles.}
\label{fig1}
\end{figure*}

In this work, we report the creation, detection, and rearrangement of an array of optically levitated nanoparticles in vacuum. Ultracold atoms and molecules in an array of optical tweezers in vacuum have provide a versatile platform for large scale quantum simulation and quantum computing~\cite{Antoine30,Lukin31,Matthew2019,Kaufman32,Anderegg2019}. There are also significant progress in optical manipulation of multiple microparticles and nanoparticles in a liquid~\cite{Masuhara33,David36,Dholakia37}.
However, creating a reconfigurable array of optically levitated nanoparticles in vacuum is still challenging due to the difficulty of  loading multiple optical tweezers simultaneously and the lack of damping in vacuum to stabilize the system.  Here we employ a high NA objective lens to tightly focus  laser beams in the vertical direction against the gravity to create an array of optical tweezers. In this layout, the tightly focused optical tweezers  provide large gradient forces for trapping and the gravity helps compensate the scattering force and photophoretic force to stabilize the system. As a result, our optical tweezers array can levitate silica nanoparticles at low pressures without feedback cooling.
In addition, we use an independent detection laser beam to monitor the motion of each nanoparticle in the array. A control laser beam is used to move and rearrange nanoparticles to desired  patterns. As an application of this technique, we choose two nanoparticles  without  rotational signals in an array to synthesize a silica nanodumbbell in-situ, and drive the created nanodumbbell to rotate at GHz frequencies. On-demand assembly of optically levitated nanoparticle arrays in vacuum will be important for creating macroscopic quantum entanglement~\cite{Vitali11,Zhang12,Hendrik13,Oriol20} and studying complex phases of interacting systems~\cite{Kishan38,Svak42,Vijayan2022,Rieser2022,liu2020}.

\begin{figure*}[t]
\centering
\centerline{\includegraphics[width=0.9\linewidth]{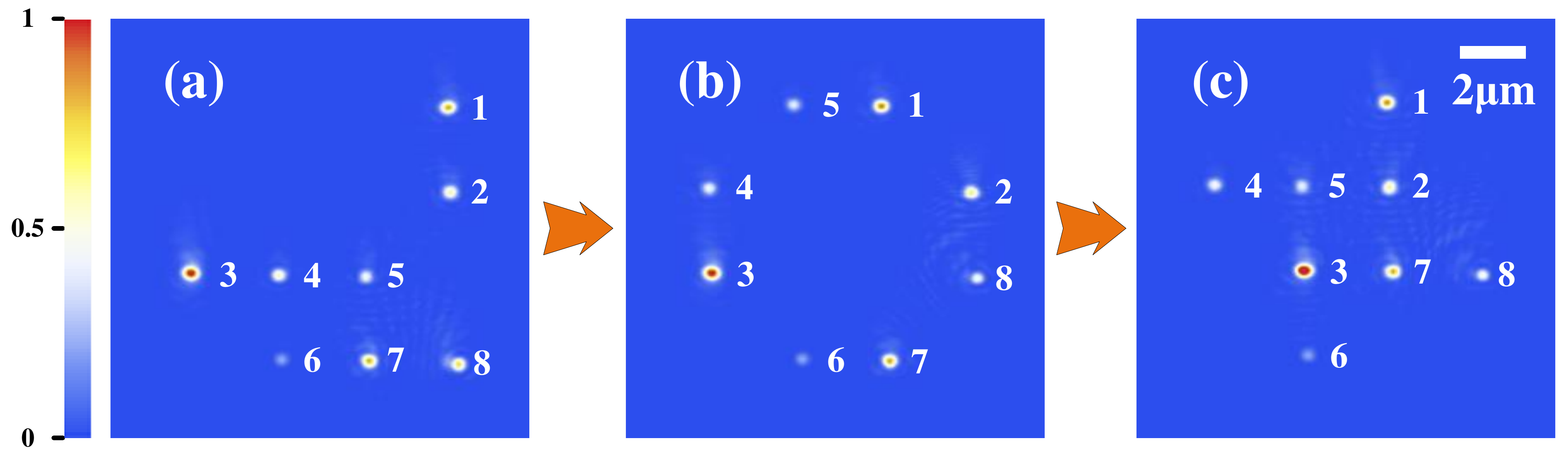}}
\caption{ \textbf{Rearranging the pattern of a nanoparticle array.} (a) The initial pattern of  8 levitated nanoparticles in a $4\times4 $  array of optical tweezers. (b)-(c) Rearranged patterns of the nanoparticle array. } \label{fig2}
\end{figure*}

The experimental protocol is illustrated in Fig. 1(a). A two-dimensional (2D) array of 1064 nm laser beams is created by a pair of orthogonal acousto-optic deflectors (AOD) driven by a multitone radio-frequency (RF) signal. The resulting beams are focused by a NA=0.95 objective lens to create an array of optical tweezers in a vacuum chamber. The optical tweezers propagate along vertical direction against the gravity. The power of each trapping beam is about 200 mW. More details can be found in Methods. Fig. 1(b) shows a simulated trapping potential of a $3 \times 3$  array of optical tweezers for single 170~nm nanoparticles. An auxiliary 1064 nm laser controlled by a motor-driven reflective mirror is used to rearrange  nanoparticles into arbitrary patterns on demand. In addition, we use a 532 nm detection beam to measure the motion of trapped nanoparticles and a 532 nm illuminating beam for imaging.  An photo of 9 nanoparticles trapped in an $3\times3 $  array is shown in Fig. 1(c). The horizontal distance of two adjacent columns is 1.77 $\mu$m and the vertical distance of two adjacent rows is 2.66 $\mu$m.

\begin{figure*}[t]
\centering
\centerline{\includegraphics[width=0.9\linewidth]{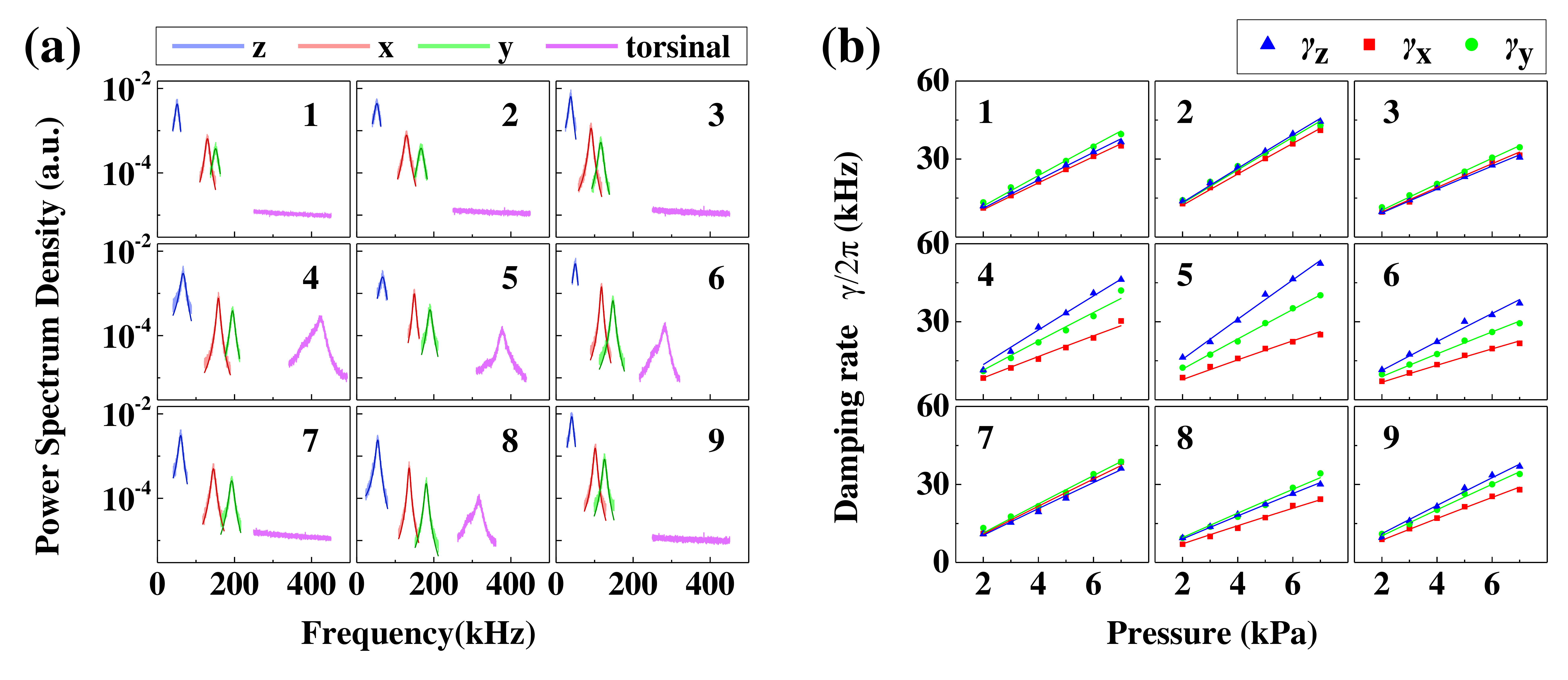}}
\caption{\textbf{Characterization of each nanoparticle in an optically-levitated nanoparticle array.} (a) The power spectra of the CoM and torsional motions for the $3\times3 $ trapped nanoparticles as shown in Fig. 1(c) at 2000 Pa. The blue, red, and green  traces correspond to the power spectra of the CoM motions along the $z$ axis, $x$ axis, and $y$ axis. The pink traces are the power spectra of torsional motions. (b) The damping rates of the CoM motions for the 9 trapped nanoparticles as a function of the pressure. The dots are measured data and the solid lines are linear fittings. }
\label{fig3}
\end{figure*}

We can use the auxiliary trapping beam to rearrange the
pattern of an array. By controlling the orientation of this
laser, we can transport and remove the nanoparticles at the near atmospheric pressures with almost perfect success probabilities. Fig. 2 shows an example of such on-demand rearrangement. First, we load the nanoparticles into a $4\times4 $  array of optical tweezers as shown in Fig. 2(a). We can see the initial filling of the array is probabilistic. Then we perform the rearrangement procedure to construct arbitrary and flexible patterns as shown in Fig. 2(b) and (c).

\begin{figure*}[t]
\centering
\centerline{\includegraphics[width=\linewidth]{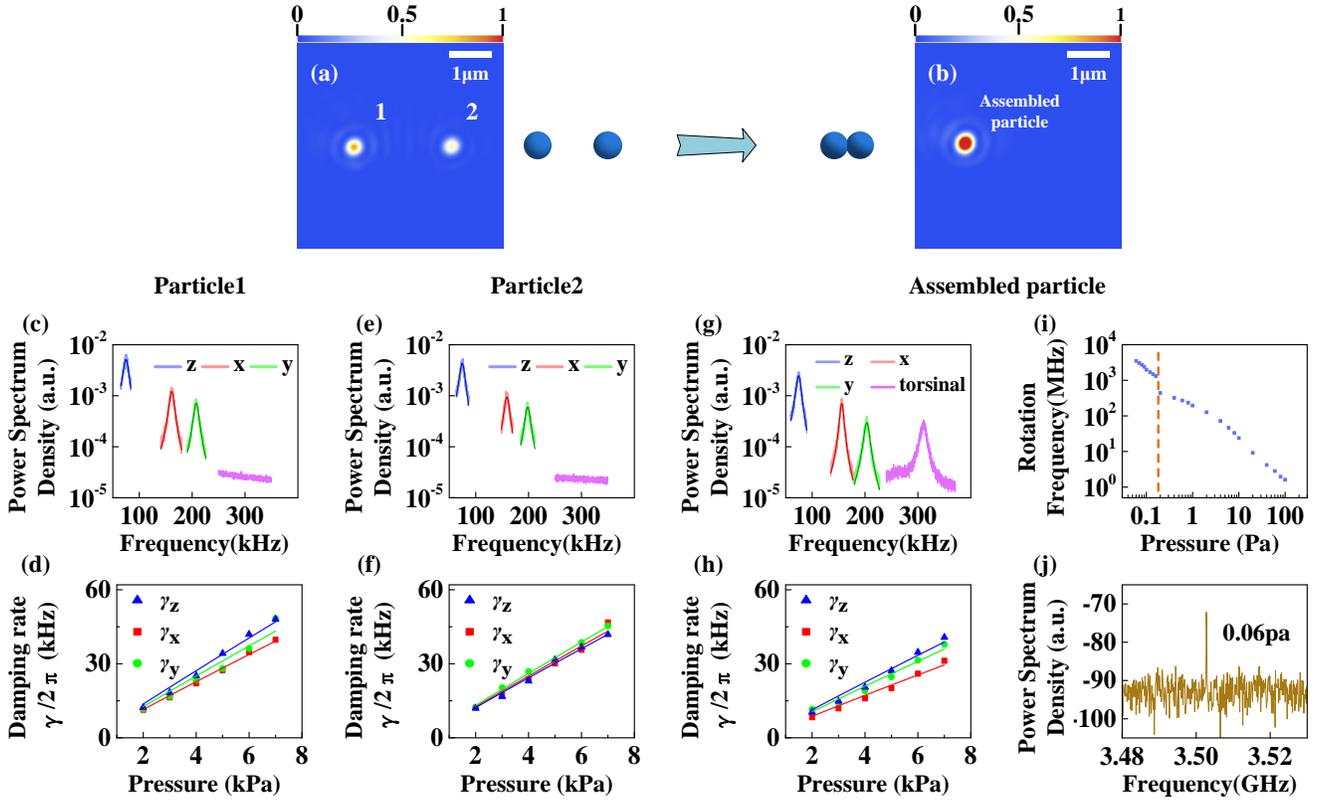}}
\caption{\textbf{In-situ synthesis of a nanodumbbell by merging two optically levitated nanoparticles}. (a) An image of two separated nanoparticles in two optical tweezers. (b) An image of the assembled nanoparticle. (c), (d) and (e), (f) CoM motion signals and the corresponding damping rates as functions of the pressures for nanoparticle 1 and 2 respectively. (g) and (h) CoM motion signals and the corresponding damping rates as a function of the pressure for the assembled nanodumbbell. For (c)-(h), the trapping laser beams are linearly polarized. (i) Rotational frequency of the nanodumbbell as a function of the pressure. (j) Rotational signal of the assembled nanodumbbell at 0.06 Pa. For (i) and (j), the trapping laser beam is circularly polarized. } \label{fig4}
\end{figure*}

To exactly characterize each optically-levitated nanoparticle in the array, we use a probe beam to measure CoM and torsional motion of each nanoparticle. The CoM and torsional motion signals for each nanoparticle in an $3\times3$ array [Fig. 1(c)]  at 2000 Pa are shown in Fig. 3(a). We  find that nanoparticles labeled as ``1'', ``2'', ``3'', ``7'', and ``9''  have no signal of torsional motion, which implicates these nanoparticles are near spherical. However, the nanoparticles labeled as ``4'', ``5'', ``6'', and ``8'' display torsional motion signals, indicating anisotropy in their shapes.  Note that a linearly polarized trapping laser is used here to generate torsional motion of the trapped nanoparticle. We can model the nanoparticle as an ellipsoid with different sizes ($r_{1}>r_{2}>r_{3}$) along three major axes. The longest axis with size $r_{1}$ of the nanoparticle will tend to align with the electric field of the linearly polarized laser, which is defined as the $x$ axis. This is because the polarizability of the nanoparticle along its longest size axis is the largest~\cite{Li26}. The trapping frequencies of the CoM motion depend on the polarization of the trapping laser ~\cite{Jin43}.  The smaller radial trapping frequency is along the electric field of the linear polarized laser ($x$ axis).

The damping rate $\gamma_{i}$ of the CoM motion for a nanoparticle depends on the air pressure $p$ and its shape ($r_{i}$, $i\in\{x,y,z\}$). It is proportional to $p$ at low pressures when the mean free path of air molecules is much larger than the size of the particle.  A larger size in one direction will lead to a smaller damping rate of the CoM motion along that direction~\cite{Jin41,Li26}. Therefore, we can estimate the shape of the nanoparticle via the measured damping rates.
Fig. 3(b) shows the damping rates of the CoM motion along three orthogonal directions as a function of the pressure. When the damping rates of the CoM motion along three orthogonal directions are almost the same (the curves labeled  ``1'', ``2'', ``3'', ``7'', and ``9'' in Fig. 3(b)), we can infer that the shape of the nanoparticle is near spherical, which confirms the observation of  no torsional signal for these nanoparticles. In contrast, the damping rate ratios $\gamma_{y}/\gamma_{x}$ and $\gamma_{z}/\gamma_{x}$ of  CoM motion are large  for ``4'', ``5'', ``6'', and ``8'' nanoparticles, which means these nanoparticles are anisotropic. This agrees with the observation of torsional motion for these nanoparticles in a linearly polarized laser.

After we obtain an optically-levitated nanoparticle array, we select two nanoparticles without torsional signals from the array to synthesize a nanodumbbell that supports  fast rotation. Fig. 4(a) and (b) illustrate the assembling process. First we choose two nanoparticles from an array, which have no torsional motion signals as shown in Fig. 4(c)-(f). Then we control the auxiliary trapping beam to move one of two nanoparticles into the same trap. Two nanoparticles will stick together to become a nanodumbbell with 25$\%$ probability of success in a single trap. There are several other different situations for the two nanoparticles merged to a single trap. For example, the nanoparticles may be lost from the trap. The two nanoparticles may also remain separated in a trap, which may be due to the repulsive Coulomb force between them if they have the same sign of charges. These situations are discussed further in the supplementary material.

Figure 4(b) shows the image of the assembled nanodumbbell, whose intensity is larger than the individual nanoparticles. The CoM motion signals and the torsional motion signal are measured simultaneously, as shown in Fig. 4(g). There is almost no change in the trapping frequencies of the CoM motion after assembly. The damping rates of the assembled nanodumbbell at different pressures are shown in Fig. 4(h). It shows a larger difference in damping rates for the CoM motion along three orthogonal directions and illustrates the anisotropic shape of the nanodumbbell. When we change the polarization of the trapping laser from linear to circular, the nanodumbbell is driven to rotate. The rotation frequency as a function of the pressure is shown in Fig. 4(i). A maximum rotation frequency of about 1.75 GHz at 0.06 Pa is observed (Fig. 4(j)). By further feedback cooling of CoM, the higher rotation frequency can be reached~\cite{Jin28}

In conclusion, we have experimentally realized a 2D array of optically levitated nanoparticles in vacuum. The initial loading of the array is probabilistic, whereas the rearrangement procedure allows us to create defect-free arrays with high fidelity and construct flexible nanoparticle patterns on demand. By measuring the  motion information, we can characterize each trapped nanoparticle, especially the anisotropic shape of the nanoparticle. As a solid application, we choose two nanoparticles without the rotational signals from the array  to synthesize a nanodumbbell  by moving the two nanoparticles into a trap. This work opens up a variety of opportunities, ranging from the assembly of complex systems with different types of nanoparticles to precision measurements~\cite{John7,Carney_2021}. By applying cooling techniques~\cite{Vijayan2022} and utilizing optical binding~\cite{Kishan38,Svak42,Rieser2022}, this system can  be used to explore the many-body macroscopic quantum physics with interacting nanoparticles~\cite{Vitali11,Zhang12,Hendrik13,Oriol20,liu2020}.

\bibliography{reference}

\vspace{1cm}

\section*{Methods}
Our scheme to create an optically levitated nanoparticle array in vacuum is shown in Fig. 1 in the main text.  A 2D laser beam array is created by passing a 1064 nm laser through a pair of orthogonal acousto-optic deflectors (AOD) driven by a multitone radio-frequency (RF) signal. The resulting beam array is imaged with a 1:1 telescope onto a high NA (NA=0.95) objective lens, which creates an array of tightly focused optical tweezers in a vacuum chamber. An auxiliary 1064 nm laser with the orthogonal linear polarization is combined with the trapping beam array with a polarizing beam splitter (PBS1) and focused by the same objective. The orientation of this beam is controlled by a motor-driven reflective mirror to rearrange nanoparticles trapped by the optical tweezers array into any patterns. In order to obtain motion information of the trapped nanoparticles, a 532 nm probe beam is combined with the trapping beams by a dichroic mirror (DM1) before the vacuum chamber. The orientation of this 532 nm beam can be adjusted to make it focus on an arbitrary particle in the array. The strongly focused 1064 nm trapping laser beams and the 532 nm probe beam are collimated by an aspherical lens with NA=0.78. The output 532 nm laser and 1064 nm laser beams  from the vacuum chamber are separated by another diachroic mirror (DM2). The 532 nm probe beam is then split into two parts for detecting the center-of-mass (CoM) motion, and rotation or torsional motion of trapped nanoparticles. When 
measure the high-speed rotation at low pressure, we directly use the 1064 nm trapping laser to detection the rotation signal.

Commercial silica nanoparticles with a nominal diameter of 170 nm are utilized. The monodisperse nanoparticles are dispersed into the vacuum chamber by an ultrasonic nebulizer and trapped by the optical tweezer array with each beam at about 200 mW. In order to image the optically-levitated nanoparticle array, another 532 nm beam is used to illuminate those particles orthogonal to the optical axis of the high NA objective lens. The scattering light is collected by the same high NA objective to form an image on a charge-coupled device (CCD). This configuration provides a dark background and a high signal-noise ratio for imaging levitated nanoparticles~\cite{JINAPL,Jin:22}. The scattering 532 nm light intensity shown in the image provides preliminary information about the size of each trapped nanoparticle. The spatial distance among the nanoparticles can be precisely measured according to the interference fringes of the scattered 532 nm lights from the particles. After the loading procedure, we start a vacuum pump to evacuate air from the vacuum chamber.

\section*{CODE AVAILABILITY}
The custom codes that support the findings of this study are available from the corresponding author upon reasonable request.

\section*{ACKNOWLEDGMENTS}
This research was supported in part by National Natural Science Foundation of China (NSFC) (Grant No. 61975101, 11234008, 11361161002, 61571276).

\section*{AUTHOR CONTRIBUTIONS}
J.Z. designed and supervised the project. J.Y., X.Y. and J.Z. performed the experiments. All authors analysed the data and discussed the results. J.Z., J.Y., and T.L. wrote the manuscript. All authors interpreted the results and reviewed the manuscript.

\end{document}